\documentclass[a4paper, 11pt]{article}

\usepackage[dvips]{graphicx}
\usepackage{epsf}
\usepackage{amssymb}
\usepackage{a4wide}

\title{Perturbation analysis of a nonlinear equation arising \\ in the Schaefer-Schwartz model of interest rates}

\author{Be\'ata Stehl\'ikov\'a }

\date{}

\begin{document}

\maketitle

 \begin{abstract}
We deal with the interest rate model proposed by Schaefer and Schwartz, which models the long rate and the spread, defined as the difference between the short and the long rates. The approximate analytical formula for the bond prices suggested by the authors  requires a computation of a certain constant, defined via a nonlinear equation and an integral of a solution to a system of ordinary differential equations. In this paper we use perturbation methods to compute this constant. Coefficients of its expansion are given in a closed form and can be constructed to arbitrary order. However, our numerical results show that a very good accuracy is achieved already after using a small number of terms.
 \end{abstract}

\section{Introduction}

Interest rate is a rate charged for the use of the money. Its stochastic character, which can be observed on  financial markets, motivates stochastic modelling of interest rates. One class of such models, so called short rate models, is based on specifying a stochastic differential equation for the instantaneous interest rate (also called short rate) or several factors which determine it. Interest rates with other maturities are then computed from bond prices which are obtained as solutions to a parabolic partial differential equations. More details about interest rate modelling can be found, for example, in  books \cite{kwok}, \cite{brigo-mercurio}.

Multi-factor models include various factors to describe the short rate better: they consider stochastic volatility (see a book \cite{stoch-vol} for a recent treatment of these models), effect of interest rates in a monetary union to a country before it joins the union in \cite{corzo-schwartz}, \cite{zikova-stehlikova}, writing the short rate as a sum of factors for example in \cite{babbs} where the factors are interpreted as current effects of streams of economic "news". 

One of the possible approaches to calibrating a model is comparing the market interest rates to those determined by the model, see for example \cite{sevcovic-csajkova} for such a calibration of one-factor models and \cite{corzo-schwartz} for a two-factor model. This  requires s a quick computation of bond prices, since these need to be computed for all the data used in the calibration and every parameter set considered in the optimization process. Closed form solution to the partial differential equation for the bond prices is, however, available only for a few models. Therefore, there are papers finding approximate analytical solutions for more complicated models, e.g. \cite{choi-wirjanto}, \cite{stehlikova-sevcovic}, \cite{stehlikova}, \cite{honda}, \cite{stehlikova-capriotti}, 
\cite{funahashi}.

In this paper we consider a model which uses short rate and long rate and is parametrized by means of taking the long rate and the spread (which is the difference between the short and the long rates). The original model was formulated in \cite{schaefer-schwartz} in the 80s. However, modelling spread is an interesting feature also later, cf. \cite{crisis-1} and a recent paper \cite{crisis-2} for its application for predictive purposes. A generalization of the original paper \cite{schaefer-schwartz}, considering a more general form of volatilities and correlations, has been studied using econometric time series techniques in \cite{christiansen}. Model with these variables is one of the two-factor models considered in \cite{2-factor-estimates}, to which they applied their estimation methodology. We consider pricing bonds in this model and apply perturbation analysis (cf. \cite{hinch}, \cite{holmes}) to an integral and a nonlinear equation which needs to be solved in order to apply the approximate analytical formula for bond prices proposed in \cite{choi-wirjanto}.

The paper is organized as follows:
In the following section we present the Schaefer-Schwartz model \cite{schaefer-schwartz}  and the mathematical problem which is solved in evaluating their approximate formula for bond prices. In Section 3 we introduce a small parameter which is used in asymptotic expansions in subsequent subsections. Section 4 provides numerical examples of the results obtained by the expansions and compare them with values obtained numerically. The paper ends with a conclusion which summarizes the results.

\section{The Schaefer-Schwartz model and their approximate analytical solution for bond prices}

The model proposed by Schaefer and Schwartz in \cite{schaefer-schwartz} is formulated in terms of stochastic differential equations for the consol rate (defined as the yield of a bond that has a constant continuous coupon and infinite maturity, also called long rate) $\ell$ and spread $s$ between the short rate and the consol rate. The authors assume the variables to follow the system of stochastic differential equations
\begin{eqnarray}
\textrm{d}s &=& m(\mu-s) \, \textrm{d}t + \gamma \, \textrm{d}w_1, \label{eq:sde-spread} \\
\textrm{d}\ell &=& \beta(s,\ell,t) \, \textrm{d}t + \sigma \sqrt{\ell} \, \textrm{d}w_2,  \label{eq:sde-consol}
\end{eqnarray}
where $w_1, w_2$ are Wiener processed which are assumed to be uncorrelated; Schaefer and Schwartz provide references to empirical studies supporting this assumption. This specification assumes that the spread evolves according to Ornstein-Uhlenbeck process (\ref{eq:sde-spread}) which is a mean-reverting process with a conditional normal distribution. This is a reasonable assumption for the spread between the interest rates which can take both positive and negative values. The interest rates themselves should not attain negative values which motivates the square root type of volatility in the process (\ref{eq:sde-consol}) for the consol rate. Note that if it is assumed to be mean-reverting too, we obtain the well known square root Bessel process. However, as shown in \cite{schaefer-schwartz}, the function $\beta(s,\ell,t)$ have no effect on the partial differential equation for the bond prices and hence it is left in the general form.

The price $V(s,\ell,\tau)$ of a discount bond with maturity $\tau$ when the current spread equals $s$ and consol rate equals $\ell$ is a solution to the partial differential equation, see \cite[Equation (4)]{schaefer-schwartz},
\begin{equation}
\frac{1}{2}\gamma^2 \frac{\partial^2 V}{\partial s^2} + \frac{1}{2}\sigma^2 \ell \frac{\partial^2 V}{\partial \ell^2}  + 
m(\hat{\mu}-s) \frac{\partial V}{\partial s} + (\sigma^2 - \ell s) \frac{\partial V}{\partial \ell} - (\ell + s)  V - 
\frac{\partial V}{\partial \tau}  = 0 \label{eq:exact}
\end{equation}
for $\tau \in [0, T), s \in \mathbb{R}, \ell \in \mathbb{R}^{+}$, with initial condition $V(s,\ell,0)=1$, where 
\begin{equation}
\hat{\mu} = \mu - \frac{\lambda \gamma}{m} \label{eq:mu_hat}
\end{equation}
and $\lambda$ is so called market price of spread risk, which is assumed to be constant.

The approximation proposed by Schaefer and Schwartz is based on finding the exact solution an equation closely related to (\ref{eq:exact}). In particular, they change the term $s$ in the coefficient of ${\partial V}/{\partial \ell}$ to a constant $\hat{s}$. Then, the solution $\bar{V}$ of the modified equation can be separated in the form $\bar{V}(s,\ell,\tau) = X(s, \tau) Y(\ell, \tau)$ where the functions $X$ and $Y$ can be expressed in the closed form.

We refer the reader to the original paper \cite{schaefer-schwartz} for the details of the motivation (which is based on expressing the stochastic differential equations in so called risk neutral equivalent measure and temporarily disregarding their stochastic part) and mathematical derivation of the problem corresponding to finding $\hat{s}$  according to the selected criterion; now we present on the formulation of the resulting problem. The final result is, that the constant $\hat{s}$ is chosen as the solution to the nonlinear equation
\begin{equation}
\bar{\ell} = \frac{l_0 \hat{s} - \sigma^2}{\hat{s}^2 \tau} \left( 1 - e^{-\hat{s} \tau} \right) + \frac{\sigma^2}{\hat{s}},
\label{eq:nonlinear}
\end{equation}
where 
\begin{equation}
\bar{\ell} = \int_0^{\tau} \ell(t) \, \textrm{d}t \label{eq:integral}
\end{equation}
and $\ell(t)$ is the solution of the system
\begin{eqnarray}
\textrm{d}s = m(\mu-s) \, \textrm{d}t, \label{eq:ode:s} \\
\textrm{d}\ell = (\sigma^2 - s \ell) \,\textrm{d}t, \label{eq:ode:l}
\end{eqnarray}
with initial conditions $s(0)=s_0$, $\ell(0)=\ell_0$ which are the values of the spread and consol rate for which the bond price is being evaluated. They compare the results of bond prices computed by this procedure with numerically solving the partial differential equation
(\ref{eq:exact}) and conclude that the approximation gives accurate results.

However, the expressions for the integral $\bar{\ell}$ defined by (\ref{eq:integral}, which are given by the authors, are  quite complicated (cf. \cite[Equation (A6)]{schaefer-schwartz}):
\begin{itemize}
\item{for $s_0 > \hat{\mu}$
\begin{equation}
\bar{\ell} = \left( \frac{e^{-V_0} V_0^{\alpha} l_0}{m \tau} + \frac{\sigma^2 \gamma(\alpha, V_0)}{m^2 \tau} \right)
 \sum_{m=0}^{\infty} \frac{V_0^{n-\alpha} - V_{\tau}^{n-\alpha}}{(n-\alpha) n!} -
\frac{\sigma^2 \Gamma(\alpha)}{m^2 \tau} \sum_{n=1}^{\infty} \frac{V_0^n - V_{\tau}^n}{n \Gamma(\alpha + n +1)} - \frac{\sigma^2}{\alpha m}, \label{eq:A6a}
\end{equation} 
}
\item{for $s_0 = \hat{\mu}$
\begin{equation}
\bar{\ell} = \frac{l_0 \hat{\mu} - \sigma^2}{\mu^2 \tau} \left( 1- e^{-\hat{\mu} \tau} \right) + \frac{\sigma^2}{\hat{\mu}}, \label{eq:A6b}
\end{equation}
} 
\item{for $s_0 < \hat{\mu}$
\begin{eqnarray}
\bar{\ell} &=& \frac{\sigma^2}{m^2 \tau} \left[  \left( \frac{e^{V_0} V_0^{\alpha} l_ m}{\sigma^2} + \frac{V_0^{\alpha}}{\alpha} + \sum_{n=1}^{\infty} \frac{V_0^{\alpha + n}}{(\alpha + n) n!} \right) \right. \nonumber \\
&& \times \left. \left( \frac{V_{\tau}^{-\alpha} - V_0^{-\alpha}}{\alpha} + \sum_{n=1}^{\infty} \frac{(-1)^n (V_0^{n-\alpha} - V_{\tau}^{n-\alpha})}{(n-\alpha) n!} \right) 
 - \frac{1}{\alpha}  \left( m \tau + \sum_{n=1}^{\infty} \frac{(-1)^n (V_0^n - V_{\tau}^n}{n \, n!} \right) \right. \nonumber \\
&& \left. -\left( \frac{ e^{-V_{\tau}} - e^{-V_0}}{\alpha+1} + \sum_{n=2}^{\infty} \frac{\gamma(n,V_0) - \gamma(n,V_{\tau})}{(\alpha + n) n!} \right) \right], \label{eq:A6c}
\end{eqnarray}
} 
\end{itemize}
where $\Gamma(x)$ is the gamma function, $\gamma(p,x)$ is the incomplete gamma function, 
$$V_0 = \frac{|s_0 - \mu|}{m}, V_{\tau} = V_0 e^{-m \tau}, \alpha = - \frac{\mu}{m}$$
and $\alpha$ is assumed to be positive (this means $\mu<0$ which is the empirically relevant case, for which in the equilibrium the consol rate is higher as the instantaneous rate).

These complicated expressions motivate us to use perturbation methods and derive the approximations to integral (\ref{eq:integral}) as well as to the solution to the nonlinear equation (\ref{eq:nonlinear}) which are easily evaluated.

Alternatively, the system of ordinary differential equations (\ref{eq:ode:s}), (\ref{eq:ode:l}) can be easily solved numerically; 
we use this approach to obtain numerical values with which we will compare our approximations.
 
\section{Perturbation methods approach to finding $\hat{s}$}

Firstly, we note that the equation (\ref{eq:ode:s}) with initial condition $s(0)=s_0$ has the solution
$s(t) = \hat{\mu} + (s_0 - \hat{\mu}) e^{-m t}.$
We introduce the small parameter
\begin{equation}
\varepsilon = s_0 - \hat{\mu}, \label{eq:epsilon}
\end{equation}
which will be used in the subsequent computations. Note that is represents the difference between the current value of the spread and its equilibrium value. Interest rates are expressed as decimal numbers in term structure models (for example, 0.01 corresponds to the interest rate of 1 percent \textit{per annum}),  so we can expect this parameter to be "small".  Using this parameter, we express $s(t)$ as
\begin{equation}
s(t) = \hat{\mu} + \varepsilon e^{-m t}  . \label{eq:s:solution}
\end{equation}

\subsection{Asymptotic expansion of   $\ell(t)$}
We write $\ell(t)$ in the form of an asymptotic series in integer powers of $\varepsilon$:
\begin{equation}
\ell(t) = c_0(t) + c_1(t) \varepsilon + c_2(t) \varepsilon^2 + \dots  \label{eq:l:expansion}
\end{equation}
From the initial condition $\ell(0)=\ell_0$ we obtain
\begin{equation}
c_0(0)=\ell_0, \,\, c_k(0)=0 \textrm{ for } k=1,2,\dots  \label{eq:ck:initial}
\end{equation}
Substituting (\ref{eq:l:expansion}) and (\ref{eq:s:solution}) into the ordinary differential equation (\ref{eq:ode:l}) we obtain the system of equations for the functions $c_k$:
\begin{equation}
\dot{c}_0(t) = \sigma^2 - \hat{\mu} c_0(t), \,\, \dot{c}_k(t) = - \hat{\mu} c_k(t) = e^{-m t} c_{k-1}(t) \textrm{ for }  k=1,2,\dots 
\label{eq:ck:ode}
\end{equation}
Thus,
\begin{equation}
c_0(t)=c_{01} + c_{02} e^{-\mu t}, \textrm{ where } c_{01} = \frac{\sigma^2}{\hat{\mu}}, c_{02} = \ell_0 - \frac{\sigma^2}{\hat{\mu}}
\end{equation}
and the functions $c_k$ can be computed in the closed form by evaluating recursively 
$$c_k(t) = - e^{-\hat{\mu} t} \int_0^t e^{(\hat{\mu} - m) s} c_{k-1}(s) \, \textrm{d}s.$$
They turn out to be linear combinations of exponential functions; we list some of the first ones:
\begin{eqnarray}
c_1(t)&=& c_{11} e^{-(\hat{\mu} +m)t} + c_{12} e^{-mt} + c_{13} e^{- \hat{\mu}t}, \nonumber \\
c_2(t)&=& c_{21} e^{-(\hat{\mu} + m )t} +c_{22} e^{-(\hat{\mu} + 2m )t} + c_{23} e^{-2 m t} +c_{24}  e^{- \hat{\mu}t}, \nonumber \\
c_3(t)&=&c_{31}  e^{-(\hat{\mu} + m )t}  + c_{32}  e^{-(\hat{\mu} + 2 m )t}  + c_{33}  e^{-(\hat{\mu} + 3 m )t}  + c_{34} e^{-3 m t} + c_{35} e^{- \hat{\mu}t}, \nonumber
\end{eqnarray}
with the coefficients given by
$$c_{11}=\frac{c_{02}}{m}, c_{12}=-\frac{c_{01}}{\hat{\mu}-m}, c_{13} = -(c_{11}+c_{12}),$$
$$c_{21} = \frac{c_{13}}{m}, c_{22}=\frac{c_{11}}{2m}, c_{23} = -\frac{c_{12}}{\hat{\mu} - 2m}, c_{24} = -(c_{21}+c_{22}+c_{23}),$$
$$c_{31} = \frac{c_{24}}{m},  c_{32}=\frac{c_{21}}{2 m}, c_{33} =\frac{c_{22}}{3 m}, c_{34} = - \frac{c_{34}}{\hat{\mu} - 3m}, 
c_{35} = -(c_{31}+c_{32} +c_{33}+c_{34}).$$
We assume that $\hat{\mu}$ is not an integral multiple of $m$ which is satisfied in a generic case since $\hat{\mu}$ is the adjusted equilibrium value of the spread (in financial setting, its equilibrium value under the equivalent risk-neutral pricing measure) given by 
(\ref{eq:mu_hat}) and $m$ is the speed of mean-reversion of spread to the equilibrium value.

\subsection{Expansion of the integral ${\bar{\ell}}$}

Since the coefficients $c_k(t)$ of the expansion (\ref{eq:l:expansion}) are linear combinations of exponentials, we can easily derive a similar expansion for the integral (\ref{eq:integral}). We write (since it is this multiple of the integral that will be used in the following section)
\begin{equation}
\tau \bar{\ell}(\tau) = L_0(\tau) + L_1(\tau) \varepsilon + L_2(\tau) \varepsilon^2 + \dots
\label{eq:tau_bar_expansion}
\end{equation}
Then, $L_k (\tau) = \int_0^{\tau} c_k(t) \, \textrm{d}t$ and the first terms are given by
\begin{eqnarray}
L_0(\tau) &=& L_{01} \tau + L_{02} e^{-\hat{\mu} \tau} + L_{03}, \nonumber \\
L_1(\tau) &=& L_{11} e^{-(\hat{\mu} + m)\tau} + L_{12} e^{-m \tau} + L_{13} e^{-\hat{\mu} \tau} + L_{14}, \nonumber \\
L_2(\tau) &=& L_{21} e^{-(\hat{\mu} + m)\tau} + L_{22} e^{-(\hat{\mu} + 2 m)\tau} + L_{23} e^{-2 m \tau} + L_{24} e^{-\hat{\mu} \tau} + L_{25}, \nonumber \\
L_3(\tau) &=& L_{31} e^{-(\hat{\mu} + m)\tau} + L_{32} e^{-(\hat{\mu} + 2m)\tau} + L_{33} e^{-(\hat{\mu} + 3 m)\tau} + L_{34} e^{- 3 m \tau} + L_{35} e^{-\hat{\mu} \tau} + L_{36}, \nonumber 
\end{eqnarray}
with the coefficients given by
$$L_{01} = c_{01}, L_{02} = - \frac{c_{02}}{\hat{\mu}}, L_{03} = \frac{c_{02}}{\hat{\mu}},$$

$$L_{11} = -\frac{c_{11}}{\hat{\mu} + m}, L_{12} = -\frac{c_{12}}{m}, L_{13} = - \frac{c_{13}}{\hat{\mu}},
 L_{14} = -\sum_{i=1}^3 L_{1j} $$
% (L_{11}+L_{12}+L_{13}),$$

$$L_{21} = -\frac{c_{21}}{\hat{\mu} + m}, L_{22} = -\frac{c_{22}}{\hat{\mu} + 2m}, L_{23} = -\frac{c_{23}}{2m}, L_{24} = -\frac{c_{24}}{\hat{\mu}},
 L_{25} = -\sum_{i=1}^4 L_{2j} $$
%  L_{25} = -(L_{21}+L_{22}+L_{23}+L_{24}),$$

$$L_{31} = -\frac{c_{31}}{\hat{\mu} + m}, L_{32}=-\frac{c_{32}}{\hat{\mu} +1 m}, L_{33}=-\frac{c_{33}}{\hat{\mu} + 3m}, L_{34} = -\frac{c_{34}}{3 m}, L_{35} - \frac{c_{35}}{\hat{\mu}},
 L_{36} = -\sum_{i=1}^5 L_{3j} $$
%  L_{36} = -(L_{31}+L_{32}+L_{33}+L_{34}+L_{35}).$$

\subsection{Expansion of $\hat{s}$}

We rewrite the equation (\ref{eq:nonlinear}) into the form
\begin{equation}
\tau \bar{\ell} \hat{s}^2 = (\ell_0 \hat{s} - \sigma^2) (1-e^{-\hat{s} \tau}) + \sigma^2 \hat{s} \tau 
\label{eq:nonlinear:mod} 
\end{equation}
and write its solution $\hat{s}$ as
\begin{equation}
\hat{s} = k_0(\tau) + k_1(\tau) \varepsilon + k_2(\tau) \varepsilon^2 + \dots \label{eq:s_hat_expansion}
\end{equation}

Now, it is easy to write both the left-hand side and the right-hand side of (\ref{eq:nonlinear:mod}) in an asymptotic expansion in 
$\varepsilon$, using (\ref{eq:s_hat_expansion}), expansion of $\tau \bar{\ell}$ given by (\ref{eq:tau_bar_expansion}) and by writing (in what follows, for the sake of brevity we  omit argument  $\tau$ in the coefficients)
\begin{eqnarray}
1-e^{-\hat{s} \tau} &=& 1 - e^{-k_0 \tau} e^{- (k_1 \varepsilon + k_2 \varepsilon^2 + \dots)\tau } \nonumber \\
&=& 1 - e^{-k_0 \tau} \left[  1 - (k_1 \varepsilon + k_2 \varepsilon^2 - \dots)\tau + \frac{1}{2} 
(k_1 \varepsilon + k_2 \varepsilon^2 + \dots)^2 \tau^2  - \dots \right]. \nonumber
\end{eqnarray}

Matching the coefficients of order $O(\varepsilon^k)$ in (\ref{eq:nonlinear:mod}) we obtain that
\begin{eqnarray}
k_0 &=& \hat{\mu} \label{eq:k0} \\
\left[ (\ell_0 k_0 - \sigma^2) e^{-k_0 \tau} \tau + \ell_0 (1- e^{-k_0 \tau}) + \sigma^2 \tau - 2 L_0 k_0 \right] k_n &=& rhs_n,
\label{eq:kn}
\end{eqnarray}
where for $n=1,2,\dots$, the right-hand side $rhs_n$ of (\ref{eq:kn}) is expressed by the previously computed quantities. Again, we write the first terms which will be used in our numerical experiment:
\begin{eqnarray}
rhs_1 &=&  L_1 k_0^2, \nonumber \\
rhs_2 &=& L_0 k_1^2 + 2 L_1 k_0 k_2 + L_2 k_0^2 + \frac{1}{2} k_1^2 \tau^2 (\ell_0 k_0 - \sigma^2) e^{-k_0 \tau} - \ell_0 k_1^2 \tau e^{-k_0 \tau}, \nonumber \\
rhs_3 &=& 2 L_0 k_1 k_2 + L_1 (2 k_0 k_2 + k_1^2) + 2 L_2  k_0 k_1 + L_3 k_0^3 - (\ell_0 k_0 - \sigma^2) e^{- k_0 \tau} (-\tau^2 k_1 k_2 + \frac{1}{6} \tau^3 k_1^3) 
\nonumber \\ &&
- \ell_0 k_1 e^{-k_0 \tau} (k_2 \tau - \frac{1}{2} \tau^2 k_1^2) - \ell_0 k_2 e^{-k_0 \tau} k_1 \tau. \nonumber
\end{eqnarray}
Note that the result (\ref{eq:k0}) is in accordance with comparing (\ref{eq:A6b}) with (\ref{eq:nonlinear}) from which it follows that $\hat{s}=\hat{\mu}$ if $s_0 = \hat{\mu}$ (i.e., $\varepsilon = 0$) as it is also noted in the paper \cite{schaefer-schwartz}.

\section{Numerical results}

We use the values which are used as base parameters in \cite{schaefer-schwartz}:
$m=0.72, \mu = -0.01, \gamma=0.007, \sigma^2=0.0003, \lambda=0$.
The initial values of spread and consol rate considered in \cite{schaefer-schwartz} are $[-0.05,0.05]$ for $s_0$ and $[0,0.2]$ for $\ell_0$. We take $\ell_0=0.1$ (middle of their range) and the following three choices  of $s_0$: the  borderline cases $s_0=-0.05$ and $s_0=0.05$, and the middle of the interval $s_0=0$. These correspond to $\varepsilon=-0.04$, $\varepsilon=0.06$ and $\varepsilon=0.01$.
 
\begin{figure}[!ht]
 \centerline{
 \includegraphics[width=0.85\textwidth]{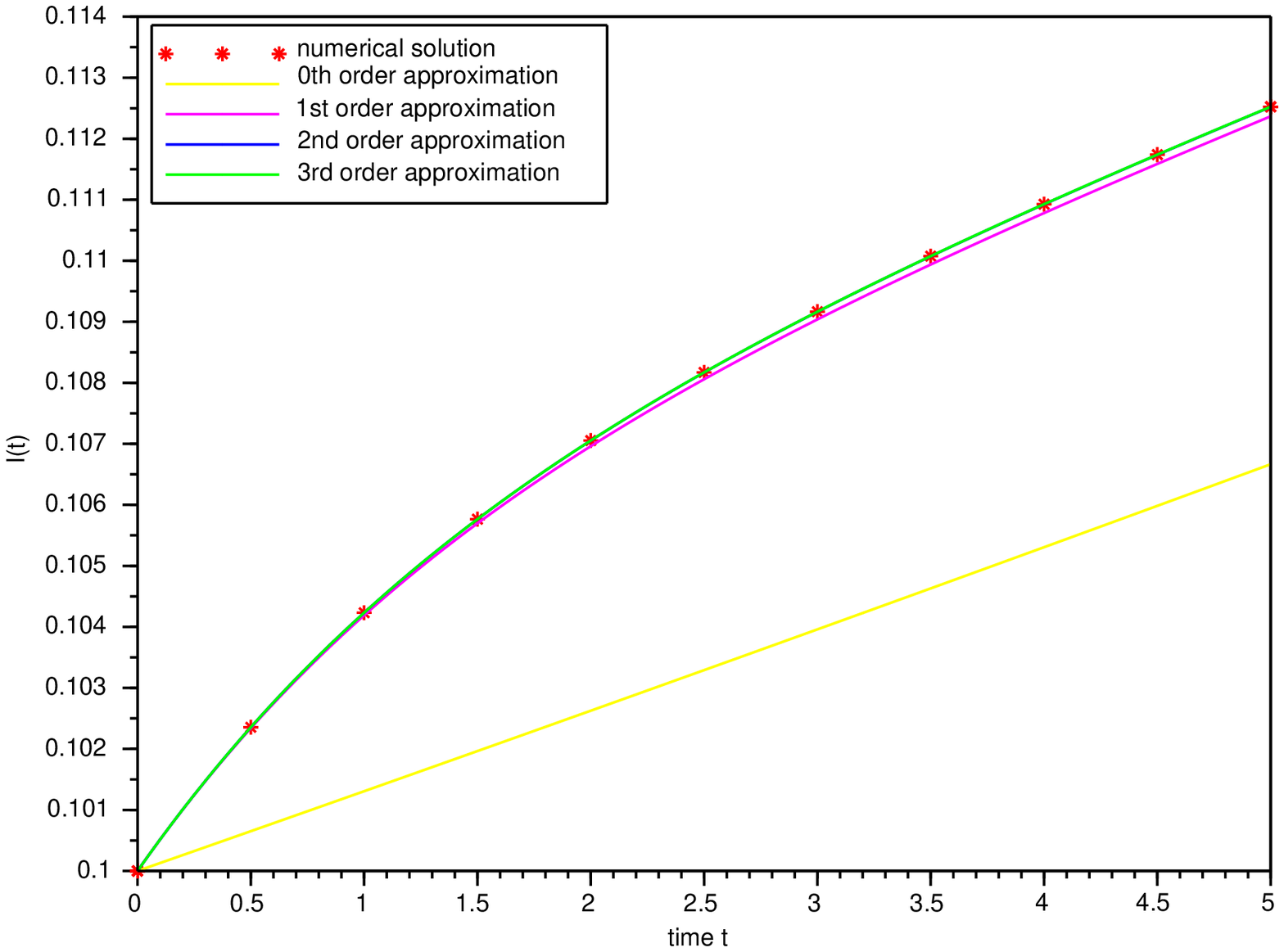} 
  }
\caption{Comparison for $s_0=-0.05$}
\label{fig:ode1}
\end{figure}

\begin{figure}[!ht]
 \centerline{
 \includegraphics[width=0.85\textwidth]{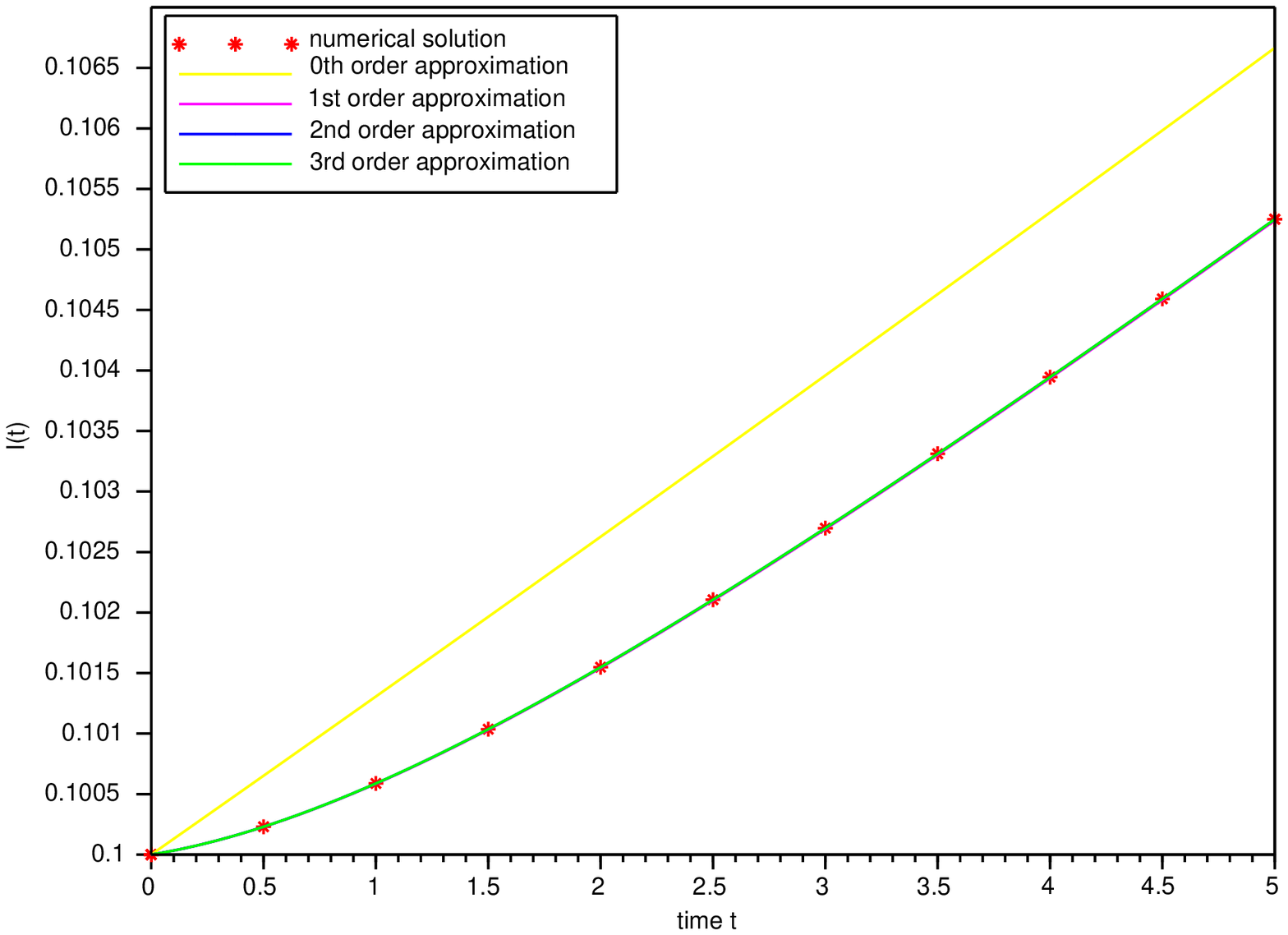} 
  }
\caption{Comparison for $s_0=0$}
\label{fig:ode2}
\end{figure}

\begin{figure}[!ht]
 \centerline{
 \includegraphics[width=0.85\textwidth]{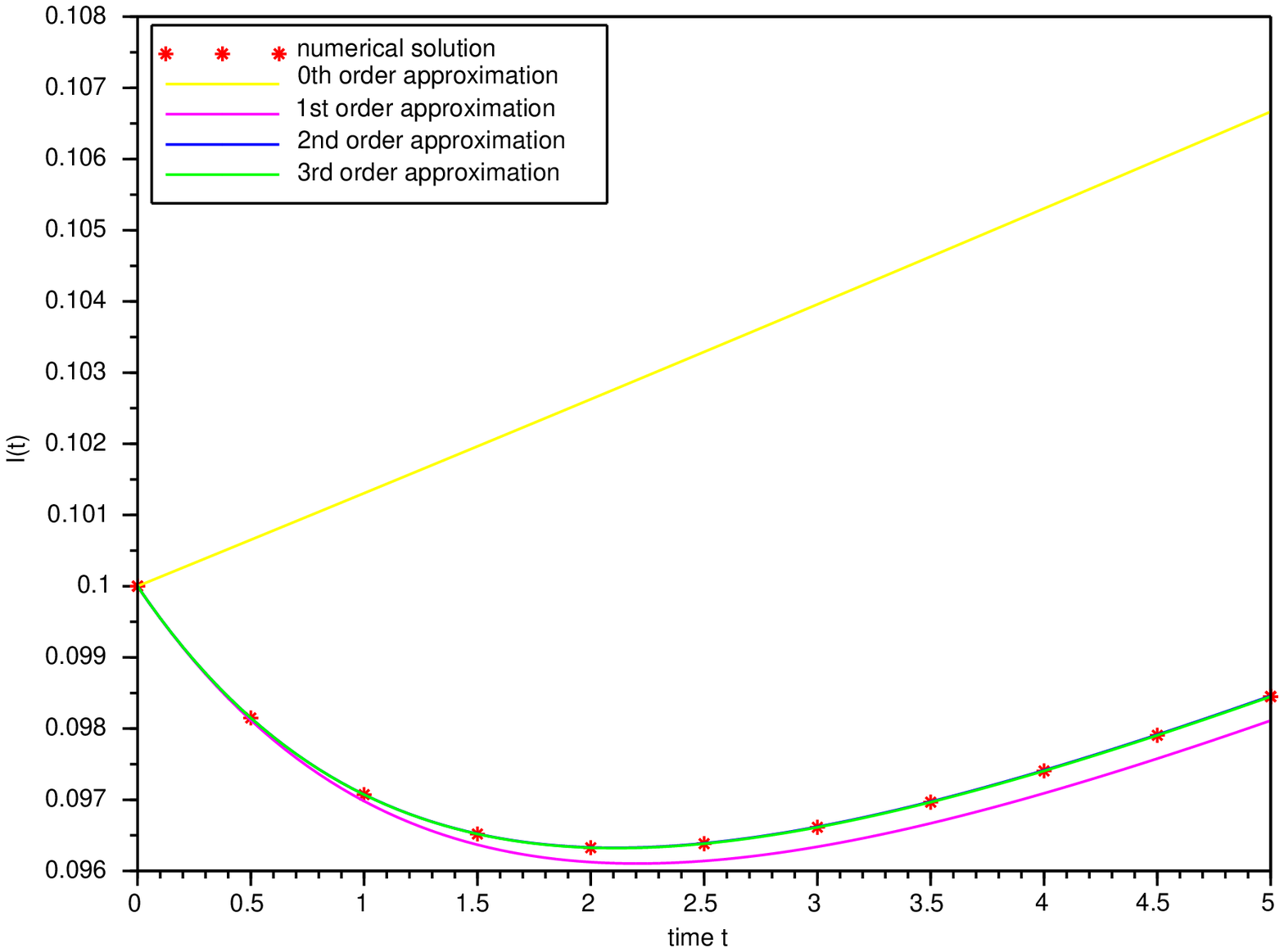} 
  }
\caption{Comparison for $s_0=0.05$}
\label{fig:ode3}
\end{figure}

Figures \ref{fig:ode1}, \ref{fig:ode2} and \ref{fig:ode3} show the comparison of successive approximations of $\ell(t)$ with the numerical solution of the ordinary differential equation (\ref{eq:ode:l}) with substituted (\ref{eq:s:solution}) for $s(t)$.
Table \ref{tab:integral} shows the successive approximations of the term $\tau \bar{\ell}$, where $\bar{\ell}$ is the integral (\ref{eq:integral}) and their comparison with the results obtained by integrating the numerical solution of (\ref{eq:ode:l}) for a bond with maturity $\tau=1$ year.
Finally, Table  \ref{tab:s_hat} shows  the successive approximations of $\hat{s}$, the quantity of interest, compared to its numerically obtained value, again for $\tau=1$.

\begin{table}[!ht]
\begin{center}
\begin{tabular}{lcccc}
\hline order &  $s_0 = -0.05$ & $s_0=0$ &  $s_0 = 0.05$ \\ 
 \hline  0 & 0.1006522   &  0.1006522   &  0.1006522  \\ 
 1 & 0.1022593  &    0.1002504  & 0.0982415  \\ 
 2 &   0.1022755  &  0.1002514   &  0.0982780  \\ 
 3 &  0.1022756   & 0.1002514   &  0.0982776 \\ 
\hline numerical value  & 0.1022756 &  0.1002514    & 0.0982776 \\ 
\end{tabular} 
\caption{ Approximations of the term $\tau \bar{\ell}$}
\label{tab:integral} 
\end{center}
\end{table}

\begin{table}[!ht]
\begin{center}
\begin{tabular}{lccc}
\hline order &  $s_0 = -0.05$ &   $s_0=0$ &  $s_0 = 0.05$ \\ 
 \hline  0 & -0.01  & -0.01  & -0.01  \\ 
 1 & -0.0418965   &   -0.0020259 &   0.0378448  \\ 
 2 &   -0.0418789  & -0.0020248  &    0.0378844   \\ 
 3 &  -0.0418789  & -0.0020248 &    0.0378844  \\ 
\hline numerical value  & -0.0418789  & -0.0020248   & 0.0378844 \\ 
\end{tabular} 
\caption{ Approximations of $\hat{s}$}
\label{tab:s_hat} 
\end{center}
\end{table}

% \section*{Acknowledgments}
% This research was supported by VEGA 1/0747/12 grant.

% \newpage

\section{Conclusions}
We proposed an application of perturbation methods to approximating the variable which is needed to evaluate the analytical approximation formula for bond prices in Schaefer and Schwartz model \cite{schaefer-schwartz}. We took the difference between the current level of spread and its equilibrium value as a small parameter and constructed expansions which finally led to an expansion of the constant needed in the approximation formula. Coefficients are given in a closed form and our numerical results show a very good approximation power, which suggests that this approach can be successfully used in practical evaluation of the model.


\begin{thebibliography}{99}

\bibitem{crisis-1}
R. Ahrens: \textit{Predicting recessions with interest rate spreads: a multicountry regime-switching analysis}.
Journal of International Money and Finance 21, 2002, pp. 519-537.

\bibitem{babbs}
S. H. Babbs, K. B. Nowman: \textit{Kalman filtering of generalized Vasicek term structure models}. 
Journal of Financial and Quantitative Analysis 34, 1999, pp. 115-130.

\bibitem{brigo-mercurio}
D. Brigo, F. Mercurio: \textit{Interest Rate Models - Theory and Practice: With Smile, Inflation and Credit}.
Springer, 2007.

\bibitem{christiansen}
Ch. Christiansen: \textit{Multivariate term structure models with level and heteroskedasticity effects}.
Journal of Banking and Finance 29, 2005, pp. 1037-1057.

\bibitem{crisis-2}
Ch. Christiansen: \textit{Predicting severe simultaneous recessions using yield
spreads as leading indicators}. Journal of International Money and Finance 32, 2013, pp. 1032-1043.

\bibitem{choi-wirjanto}
Y. Choi, T. S. Wirjanto: \textit{An analytic approximation formula for pricing zero-coupon bonds}. 
Finance Research Letters 4, 2007, pp. 116-126.

\bibitem{corzo-schwartz}
T. S. Corzo,   E. S. Schwartz: \textit{Convergence within the EU: evidence from interest rates}.  Economic Notes 29, 2000, pp. 243-266 

\bibitem{stoch-vol}
J.-P. Fouque, G. Papanicolaou, R. Sircar, K. Solna: \textit{Multiscale Stochastic Volatility for Equity, Interest Rate, and Credit Derivatives}. Cambridge University Press, 2011.

\bibitem{funahashi}
H. Funahashi, T. Fukui: \textit{A Unified Approximation Formula for Zero-Coupon Bond Prices}. Working paper, 2014. Available at SSRN: \texttt{http://ssrn.com/abstract=2472503}

\bibitem{2-factor-estimates}
L. G\'omez-Valle, J. Martin\'ez-Rodrigu\'ez: \textit{Improving the term structure of interest rates: Two-fator models}. 
Interntional Journal of Finance and Economics, 15, 2010, pp. 275-287. 

\bibitem{hinch}
E. J. Hinch: \textit{Perturbation Methods}. Cambridge University Press, 1991.

\bibitem{holmes}
M. H. Holmes: \textit{Introduction to Perturbation Methods.}  Springer, 2012.

\bibitem{honda}
T. Honda, K. Tamaki, T. Shiohama: \textit{Higher order asymptotic bond price valuation for
interest rates with non-Gaussian dependent innovations}. Finance Research Letters 7, 2009, 
pp. 60-69.

\bibitem{kwok}
Y. - K. Kwok: \textit{Mathematical Models of Financial Derivatives}. Springer, 2008.

\bibitem{schaefer-schwartz}
 S. M. Schaefer, E. S. Schwartz: \textit{A Two-Factor Model of the Term Structure: An Approximate Analytical Solution}. The Journal of Financial and Quantitative Analysis, Vol. 19, No. 4, 1984, pp. 413-424.

\bibitem{stehlikova}
B. Stehl\'ikov\'a: \textit{A simple analytic approximation formula for the bond price in the Chan-Karolyi-Longstaff-Sanders model. }
International Journal of Numerical Analysis and Modeling - Series B, Volume 4, Number 3, 2013, pp. 224-234 

\bibitem{stehlikova-capriotti}
B. Stehl\'ikov\'a, L. Capriotti: \textit{An Effective Approximation for Zero-Coupon Bonds and Arrow-Debreu Prices  in the Black-Karasinki  Model}. To appear in International Journal of Theoretical and Applied Finance.

\bibitem{stehlikova-sevcovic}
B. Stehl\'ikov\'a, D. \v Sev\v covi\v c: \textit{Approximate formulae for pricing zero-coupon bonds and their asymptotic analysis}.
International Journal of Numerical Analysis and Modeling, Volume 6, Number 2, 2009, pp. 274-283. 

\bibitem{sevcovic-csajkova}
D. \v Sev\v covi\v c, A. Urb\'anov\'a Csajkov\'a: \textit{On a two-phase minmax method for parameter estimation of the Cox, Ingersoll, and Ross interest rate model}. Central European Journal of Operation Research, 13, 2005, pp. 169-188.

\bibitem{zikova-stehlikova}
Z. Z\'ikov\'a,  B. Stehl\'ikov\'a: \textit{Convergence model of interest rates of CKLS type}. Kybernetika 3, 2012 pp. 567-586.

\end{thebibliography}
\end{document}